\documentclass[]{spie}  

\usepackage{amsmath,amsfonts,amssymb}
\usepackage{graphicx}
\usepackage[colorlinks=true, allcolors=blue]{hyperref}

\usepackage{aas_macros, gensymb, xcolor, makecell, booktabs, threeparttable}

\title{Investigating the differential limb coupling effect for diffraction-limited spectrographs with PARVI}

\author[a]{Andrea S.J. Lin}
\author[b]{Ashley D. Baker}
\author[c]{Samuel Halverson}
\author[b]{Nemanja Jovanovic}
\author[a,c]{Dimitri Mawet}
\author[d]{Rebecca Oppenheimer}
\author[c]{Garreth Ruane}
\author[c]{Gautam Vasisht}
\affil[a]{Department of Astronomy, California Institute of Technology, 1200 E California Blvd, Pasadena, CA 91125, USA}
\affil[b]{Caltech Optical Observatories, California Institute of Technology, 1200 E California Blvd, Pasadena, CA 91125, USA}
\affil[c]{Jet Propulsion Laboratory, California Institute of Technology, 4800 Oak Grove Drive, Pasadena, California 91109}
\affil[d]{American Museum of Natural History, 200 Central Park West, New York, NY, USA}

\authorinfo{Further author information: (Send correspondence to A.S.J.L. and/or A.D.B.)\\A.S.J.L: E-mail: asjlin@caltech.edu\\A.D.B: E-mail: abaker@caltech.edu}

\begin{document} 
\maketitle

\begin{abstract}
A promising new architecture for extreme-precision radial velocity (EPRV) spectrographs, hunting for small-amplitude stellar Doppler shifts induced by orbiting planets, is to build diffraction-limited instruments by using single-mode fibers fed by an adaptive optics system. However, the target stars are partially resolved when observing at the diffraction limit, and the resulting RVs are expected to be affected by differential limb coupling (DLC), an effect where the red- and blue-shifted sides of the stellar disk are coupled unequally into the spectrograph, producing an RV error term on the order of m/s for nearby EPRV target stars for the upcoming HISPEC spectrograph for Keck~II. We present our efforts to directly measure the RV shifts resulting from DLC for the first time, using the diffraction-limited spectrograph PARVI, in order to verify the expected behavior of DLC and subsequently develop mitigation strategies for HISPEC and other future instruments. We outline an observing strategy designed to produce DLC-induced RV shifts of hundreds of m/s, and describe the execution of this experiment with PARVI. We use these data to characterize the PARVI tip-tilt guide camera and its performance, and have begun analysis of the derived RVs, though this investigation has proven complicated since the RVs are deeply entangled with other instrumental and algorithmic effects.
\end{abstract}


\section{INTRODUCTION}
\label{sec:intro}  

The radial velocity (RV) method is one of the foundational methods for discovering and characterizing exoplanets, based on measuring the Doppler shift of a stellar spectrum resulting from the star's reflex motion in response to the gravitational pull of orbiting planets. Advancements in spectrograph design and technology over the past several decades have enabled the current generation of planet-hunting instruments to detect RV signals with amplitudes at the $\lesssim$~1~m/s level (e.g., Refs. \citenum{John2023MNRAS.525.1687J, Basant2025ApJ...982L...1B, SuarezMascareno2025A&A...700A..11S}), but further advances are still needed in order to reach the $<$~10~cm/s RV precision required to detect an Earth-mass planet at 1~au around a Sun-like star. This is the goal of the field of extreme-precision radial velocities (EPRV).

The 2021 EPRV Working Group report \cite{Crass2021arXiv210714291C} not only outlined the major scientific and technical challenges facing the EPRV field, but also put forth several recommendations for promising new instrument architectures and technologies to be explored. 
One such approach is to build EPRV spectrographs that operate at the diffraction limit, using single-mode fiber (SMF) inputs fed by high-performance adaptive optics (AO) systems, instead of the traditional seeing-limited, multi-mode fiber (MMF) architecture. Operating at the diffraction limit completely decouples the size of the spectrograph from that of the telescope --- otherwise, the size of the grating (and consequently, the rest of the optics) scales linearly with telescope diameter, either necessitating image slicing or resulting in very large instruments, both of which will present even greater challenges in the impending era of Extremely Large Telescopes. In addition, SMFs, by virtue of only having a single propagation mode, eliminate the need to mitigate spatial and temporal modal noise like in MMF spectrographs, and also the need for high-gain scrambling of the near and far fields due to having better output stability regardless of the input illumination.

The first diffraction-limited EPRV spectrographs are starting to come online, with the very first being PARVI, the PAlomar Radial Velocity Instrument \cite{Gibson2020JATIS...6a1002G, Gibson2022JATIS...8c8006G, Cale2023JATIS...9c8006C}, which went on-sky in 2019 with Palomar's 200-inch Hale Telescope (5.1~m; also known as the P200). PARVI is fed by the PALM-3000 AO system \cite{Dekany2013ApJ...776..130D, Meeker2020SPIE11448E..0WM} and covers the NIR $J$ and $H$ bands (1150--1770~nm) with an as-built spectral resolution of $R \sim$~60,000. It is now followed by iLocater \cite{Crepp2016SPIE.9908E..19C, Crass2022SPIE12184E..1PC}, which covers the $yJ$ wavelength range (970--1310~nm) at an impressively high $R \sim$~190,000, and recently achieved first light \cite{Crass2026SPIE...141494, Johnson2026SPIE...14149195} at the 8.4~m Large Binocular Telescope (LBT) with the facility's integrated AO system, the LBT Interferometer \cite{Crass2021MNRAS.501.2250C}. These two instruments will soon be joined by HISPEC \cite{Mawet2019BAAS...51g.134M, Konopacky2023SPIE12680E..07K} (High-resolution Infrared SPectrograph for Exoplanet Characterization), which is currently under construction for the 10~m Keck~II telescope \cite{Echeverri2026SPIE...14149204, Fitzgerald2026SPIE...1414924, Jovanovic2026SPIE...14149452, Sappey2026SPIE...14149343}. HISPEC will utilize the newly-upgraded HAKA AO system \cite{Lilley2024SPIE13097E..7BL, Gutierrez2026SPIE...14150145, Guthery2026SPIE...14150335} and will cover $yJHK$ (0.98--2.46 ${\mu}$m) at $R \sim$~100,000; its design also serves as a pathfinder for MODHIS (Multi-Objective Diffraction-limited High-Resolution Infrared Spectrograph) on the Thirty Meter Telescope \cite{Konopacky2023SPIE12680E..07K}.

\begin{figure}[tbp]
    \centering
    \includegraphics[width=0.7\linewidth]{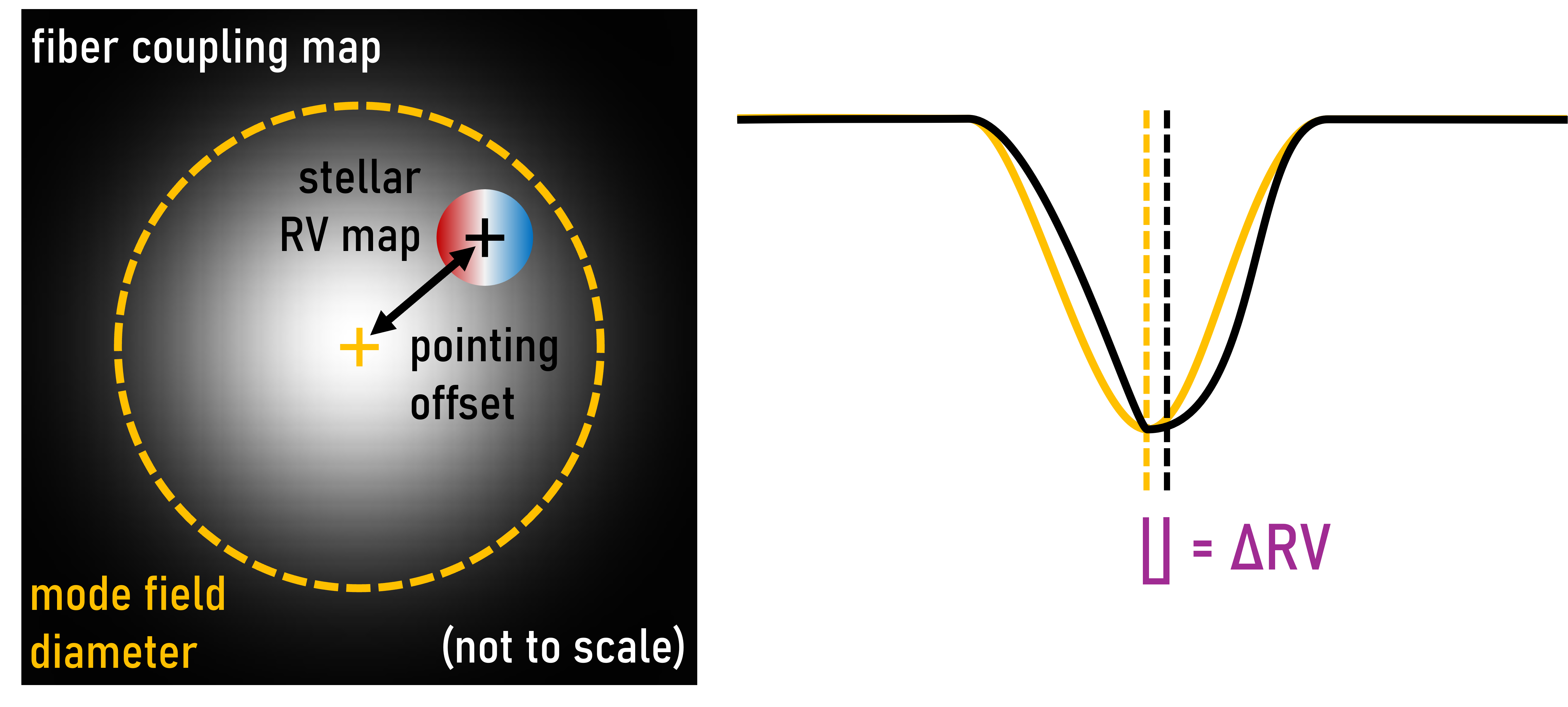}
    \caption{Schematic of the physical theory behind DLC, adapted from Ref.~\citenum{Baker2024SPIE13096E..1GB}. For a diffraction-limited instrument, the angular size of a target star can be a significant fraction of the input mode field diameter. If the star is not perfectly centered on the fiber, an unequal amount of light from the red- and blue-shifted sides of the stellar disk will be coupled into the instrument. This translates into a distortion of the spectral line profiles, which in turn causes an offset in the derived RVs when fitting for the centroids of the distorted lines. The flux imbalance, and therefore the RV shift, changes with both the direction and magnitude of the pointing offset, and thus may vary from one observation to the next.}
    \label{fig:dlc_schematic}
\end{figure}

Despite the optomechanical benefits, diffraction-limited EPRV instruments also face a potential major obstacle as a consequence of partially spatially resolving the stars they observe.
As detailed by Refs.~\citenum{Baker2024SPIE13096E..1GB} and \citenum{Guyon2024SPIE...1309719} and illustrated in Fig.~\ref{fig:dlc_schematic}, imperfect centering of the stellar PSF on the spectrograph input fiber causes the red- and blue-shifted limbs of the stellar disk to be coupled \textit{unequally} into the instrument --- hence the effect's name, differential limb coupling (DLC) --- producing non-astrophysical offsets in the measured RVs from visit to visit as a function of pointing. While the existence of DLC has been recognized for some time, and even leveraged for stellar science\footnote{Refs.~\citenum{Lesage2012SPIE.8446E..3WL} and \citenum{Lesage2014A&A...563A..86L} showed that spectral line shifts along the slit direction of a long-slit spectrograph can be used to derive the sky-projected spin axis of stars smaller than the telescope diffraction limit, and Ref.~\citenum{Walk2026SPIE...1415490} demonstrated spectro-astrometric recovery of the spin axis orientation of Humu (Altair).}, it is only recently that we have begun to consider the consequences for EPRV. Simulations by Ref.~\citenum{Baker2024SPIE13096E..1GB} found that for typical bright, nearby EPRV target stars on the Habitable Worlds Observatory precursor science list \cite{Mamajek2024arXiv240212414M} ($\alpha_* \approx$ 0.5--1.0~mas, $v \sin{i} \approx$ 2~km/s), a pointing error of merely 1~mas along the stellar equator (the direction which maximizes the impact of DLC) would result in a spurious RV shift of $\sim$~1--2~m/s for HISPEC and 8--16~m/s for MODHIS in the $yJ$ bands, completely overwhelming the rest of the instrumental RV error budget and preventing the detection of small-amplitude planetary signals.

Ref.~\citenum{Guyon2024SPIE...1309719} made the first attempts to verify the RV impacts of DLC on-sky by observing Altair with Subaru's SCExAO and a photonic lantern feeding an $R \sim$~4,000 spectrograph. (Since the $\Delta$RV from DLC scales with stellar $v \sin{i}$, as shown below in Eq.~\ref{eq:dlc}, an extremely rapid rotator like Altair is needed to produce RV shifts large enough to be measurable at $R \sim$~4,000.) They found that the resulting scatter in RVs was consistent with the predicted DLC amplitude due to SCExAO pointing jitter as measured by the photonic lantern, providing tentative, but not definitive, evidence to support our existing understanding of the RV effects of DLC.

Therefore, we set out to directly measure the $\Delta$RV shifts produced by DLC using the precision RV spectrograph PARVI, as part of our overarching goal of better quantifying DLC and developing and testing mitigation strategies before the anticipated deployment of HISPEC in 2027. In this work, we detail our efforts to measure DLC with PARVI, starting with the design of our observing experiment in Sec.~\ref{sec:expdes}, including instrument requirements, optimal target selection, and observational procedures. We then summarize the collected PARVI data in Sec.~\ref{sec:obs} and report on the current status of our analyses in Sec.~\ref{sec:analysis}.

\section{EXPERIMENTAL DESIGN}
\label{sec:expdes}

Following Ref.~\citenum{Baker2024SPIE13096E..1GB}, the RV shift resulting from DLC (at a given wavelength) scales as: 

\begin{equation}
\label{eq:dlc}
\Delta RV \propto \alpha_{\rm{offset}}\ D_{\rm{tel}}^2\ \alpha_*\ v \sin{i} = C_\lambda \left(\frac{\alpha_{\rm{offset}}}{1~\rm{mas}}\right) \left(\frac{D_{\rm{tel}}}{10~\rm{m}}\right)^2 \left(\frac{\alpha_*}{1~\rm{mas}}\right) \left(\frac{v \sin{i}}{1~\rm{km/s}}\right)
\end{equation}

\noindent where $C_{yJ} \approx$~1.0~m/s and $C_{HK} \approx$~0.2~m/s based on simulations of the HISPEC fiber coupling maps. This scaling relation motivated the design of our DLC observations as we made choices not only to maximize the potential $\Delta$RV, but also our ability to accurately measure it.

\subsection{Instrument}
\label{ssec:inst}

The ideal instrument for directly measuring DLC-induced RV shifts is not just a diffraction-limited spectrograph, but one built for precision RVs --- i.e., with high spectral resolution, allowing us to measure much smaller RV shifts (on the scale of tens or hundreds of m/s as opposed to several km/s), and an instrumental RV precision floor much lower than the expected $\Delta$RV amplitude, allowing us to be confident that the RV shifts we see are ``real'' and not simply caused by instrumental noise.

PARVI is one of the few such spectrographs in existence, and currently the only one available for routine science operations\footnote{Initially, we also attempted to use KPIC \cite{Mawet2016SPIE.9909E..0DM, Echeverri2024SPIE13096E..2DE} (Keck Planet Imager and Characterizer), which links the Keck~II AO system with the venerable NIRSPEC spectrograph \cite{McLean1998SPIE.3354..566M} at $R \sim$~30,000, but our observations were hindered by poor conditions and KPIC was subsequently decommissioned in early 2026 to allow for the HAKA AO upgrade.}. Ref.~\citenum{Cale2023JATIS...9c8006C} have shown that PARVI is capable of on-sky RV precision at the 5--10~m/s level on bright stars, and is also stable to 5--10 m/s over a night. Unfortunately, the P200 telescope is considerably smaller (5.1~m) than the 10~m Keck~II, reducing the expected $\Delta$RV by a factor of 4, but we can compensate through our choice of DLC targets. As we show below in Secs.~\ref{ssec:targets} and \ref{ssec:procedure}, the most optimal stars are expected to have $\Delta$RVs of hundreds of m/s with PARVI with our observing strategy, which should be easily measurable at the demonstrated instrument precision.

\subsection{Target selection}
\label{ssec:targets}

For our experiment, we sought to observe the stars that exhibited the largest possible $\Delta$RV shifts from DLC, so that the shifts could be robustly detected with PARVI. The two stellar terms in the DLC scaling relation (Eq.~\ref{eq:dlc}) are angular diameter ($\alpha_*$) and rotational velocity ($v \sin{i}$), so we found that the optimal targets generally fall into two categories: nearby K/M-type giants and A/F-type rapid rotators.

We compiled a list of these stars observable with PARVI (Table~\ref{tab:targets_all}) which have existing literature values of $\alpha_*$ and $v \sin{i}$, or where $\alpha_*$ can be estimated from Gaia data. We restricted $v \sin{i}$ to the range of 5--70 km/s, though we made an exception for Altair (242 km/s) in hopes of replicating the results seen with SCExAO by Ref.~\citenum{Guyon2024SPIE...1309719}. We deemed 5~km/s an approximate lower bound for confident detections of stellar rotation velocity, while above $\sim$70~km/s, the stellar features become so broad as to cover most of a PARVI spectral order, making it very difficult to derive RVs. 
Finally, we also excluded any stars which are known spectroscopic binaries or have possibly bound companions within a few arcseconds, since we do not want the DLC-induced $\Delta$RVs to be confounded with any actual orbital motion.

\begin{table}[tbp]
\caption{DLC targets observable with PARVI, sorted by expected $\Delta$RV}
\label{tab:targets_all}
\centering
\fontsize{8}{10}
\selectfont
\renewcommand{\arraystretch}{1.25}
\vspace{0.2cm}

\begin{threeparttable}
\begin{tabular}{lllcclcccc}
\toprule
\multicolumn{3}{c}{\textbf{Object}} & \multicolumn{2}{c}{\textbf{Magnitudes}} & \multicolumn{3}{c}{\textbf{Stellar Parameters}} & \multicolumn{2}{c}{\textbf{PARVI $\Delta$RV (18~mas)}} \\
\cmidrule(lr){1-3} \cmidrule(lr){4-5} \cmidrule(lr){6-8} \cmidrule(lr){9-10}
Name & RA (hms) & Dec$^{*}$ (dms) & $J$ & $K$ & SpType & $\alpha_*$ (mas) & $v \sin{i}$ (km/s) & $yJ$ (m/s) & $HK$ (m/s) \\
\midrule
Altair   & 19 50 47.00  & +08 52 05.96   & 0.35      & 0.24     & A7Vn          & 3.6       & 242   & 3270  & 653 \\
Rho Per	 & 03 05 10.59	& +38 50 24.99	 & -0.76	 & -1.92	& M4+IIIa		& 15		& 11.2	& 756	& 151 \\
Bet Peg	 & 23 03 46.45	& +28 04 58.03	 & -1.11	 & -2.37 	& M2.5II-III	& 16		& 9.7	& 698	& 140 \\
Bet Cas	 & 00 09 10.68	& +59 08 59.21	 & 1.63 	 & 1.44 	& F2III			& 2			& 71	& 639	& 128 \\
Mu Gem	 & 06 22 57.63	& +22 30 48.90	 & -0.78	 & -1.49	& M3IIIab		& 13.9		& 8.4	& 525	& 105 \\
Zet Leo	 & 10 16 41.41	& +23 25 02.34	 & 2.81		 & 2.62 	& F0IIIa		& 1.5$^{e}$	& 72	& 486	& 97 \\
Bet And	 & 01 09 43.92	& +35 37 14.00	 & -0.84	 & -1.87	& M0+IIIa		& 14		& 7.2	& 484	& 91 \\
Zet And	 & 00 47 20.32	& +24 16 01.84	 & 2.23		 & 1.56 	& K1III			& 2.5		& 39.3	& 442	& 88 \\
Ksi Gem	 & 06 45 17.36	& +12 53 44.14	 & 1.99		 & 1.68 	& F5IV-V		& 1.4		& 66	& 416	& 83 \\
Alf Cet	 & 03 02 16.77	& +04 05 23.05	 & -0.62	 & -1.67	& M1.5IIIa		& 12		& 6.9	& 373	& 75 \\
Alf Hya	 & 09 27 35.24	& --08 39 30.95	 & -0.36	 & -1.21	& K3IIIa		& 9.1		& 8.5	& 348	& 70 \\
Ups Peg	 & 23 25 22.78	& +23 24 14.76	 & 3.37		 & 3.04 	& F8III			& 1$^{e}$	& 73	& 329	& 66 \\
Del Oph	 & 16 14 20.73	& --03 41 39.57	 & -0.13	 & -1.17	& M0.5III		& 10		& 7		& 315	& 63 \\
31 Com	 & 12 51 41.92	& +27 32 26.56	 & 3.72		 & 3.35 	& G0IIIp		& 0.9$^{e}$	& 67	& 271	& 54 \\
Gam Dra	 & 17 56 36.36	& +51 29 20.02	 & -0.44	 & -1.35	& K5III			& 10		& 6		& 270	& 54 \\
Del Vir	 & 12 55 36.20	& +03 23 50.88	 & -0.11	 & -1.18	& M3+III		& 9.9		& 6		& 267	& 53 \\
Rho Gem	 & 07 29 06.71	& +31 47 04.37	 & 3.22		 & 2.97 	& F1V			& 0.85		& 59	& 226	& 45 \\
Gam Vir	 & 12 41 39.62	& --01 26 57.85	 & 2.09		 & 1.88 	& F1-F2V		& 1.6		& 30	& 216	& 43 \\
Bet UMa	 & 11 01 50.47	& +56 22 56.76	 & 2.35		 & 2.35 	& A1IV			& 1			& 47	& 212	& 42 \\
Alf Lyn	 & 09 21 03.30	& +34 23 33.21	 & 0.32		 & -0.64	& K6III			& 7.2		& 6.4	& 207	& 41 \\
Alf Cyg	 & 20 41 25.91	& +45 16 49.21	 & 0.95		 & 0.88 	& A2Ia			& 2.3		& 20	& 207	& 41 \\
Alf Cas	 & 00 40 30.44	& +56 32 14.38	 & 0.42		 & -0.25	& K0-IIIa		& 5.3		& 8.5	& 203	& 41 \\
Lam Psc	 & 23 42 02.80	& +01 46 48.15	 & 4.1		 & 4.06 	& A7V			& 0.6		& 70	& 189	& 38 \\
7 And	 & 23 12 33.00	& +49 24 22.34	 & 3.95		 & 3.77 	& F1V			& 0.6		& 62	& 167	& 33 \\
Gam Sge	 & 19 58 45.42	& +19 29 31.72	 & 0.7		 & -0.25	& M0-III		& 6			& 5.8	& 157	& 31 \\
Bet Cnc	 & 08 16 30.92	& +09 11 07.95	 & 1.06		 & 0.14 	& K4III			& 4.8		& 6.9	& 149	& 30 \\
Tet Boo	 & 14 25 11.79	& +51 51 02.67	 & 3.17		 & 2.73 	& F7V			& 1.1		& 29	& 144	& 29 \\
Eta Per	 & 02 50 41.80	& +55 53 43.77	 & 1.07		 & 0.16 	& K3-Ib-IIa		& 5.4		& 5.8	& 141	& 28 \\
Ups Gem	 & 07 35 55.35	& +26 53 44.68	 & 1.2		 & 0.24 	& M0III			& 4.8		& 5.9	& 127	& 25 \\
\bottomrule
\end{tabular}
\begin{tablenotes}
    \small
    \item $^{*}$ The actual range of declinations allowed here is $-$10\degree\ $<$ Dec $<$ +60\degree, because for our earlier PARVI observing runs, the P200 (an equatorial telescope) could not point above +60\degree\ with PARVI mounted due to mechanical restrictions.
    \item $^{e}$ These stellar diameters are estimated.
\end{tablenotes}
\end{threeparttable}
\end{table}

\subsection{Observing Procedure}
\label{ssec:procedure}

The PARVI tip-tilt stages are normally used to direct the output of the PALM-3000 AO system into the spectrograph input fiber, providing an easy method for us to deliberately offset stars from the fiber center. Based on communication with the PARVI instrument team, we are confident that the tip-tilt stages can reliably offset with accuracy on the order of 1~mas. 
Since the expected $\Delta$RV from DLC scales linearly with the offset between the center of the fiber and the center of the star (Eq.~\ref{eq:dlc}), we planned to offset in steps of $\sim$20~mas, a value chosen to be a significant fraction of the $\sim$80~mas diameter PARVI input fiber and also larger than the typical AO pointing jitter at Palomar. After deriving the plate scale of the PARVI tip-tilt guide camera (Sec.~\ref{ssec:gcam_scale}), we ultimately ended up using steps of 18~mas, since this corresponds to 0.5~pix on the guide camera.

\begin{figure}[tbp]
    \centering
    \includegraphics[width=0.4\linewidth]{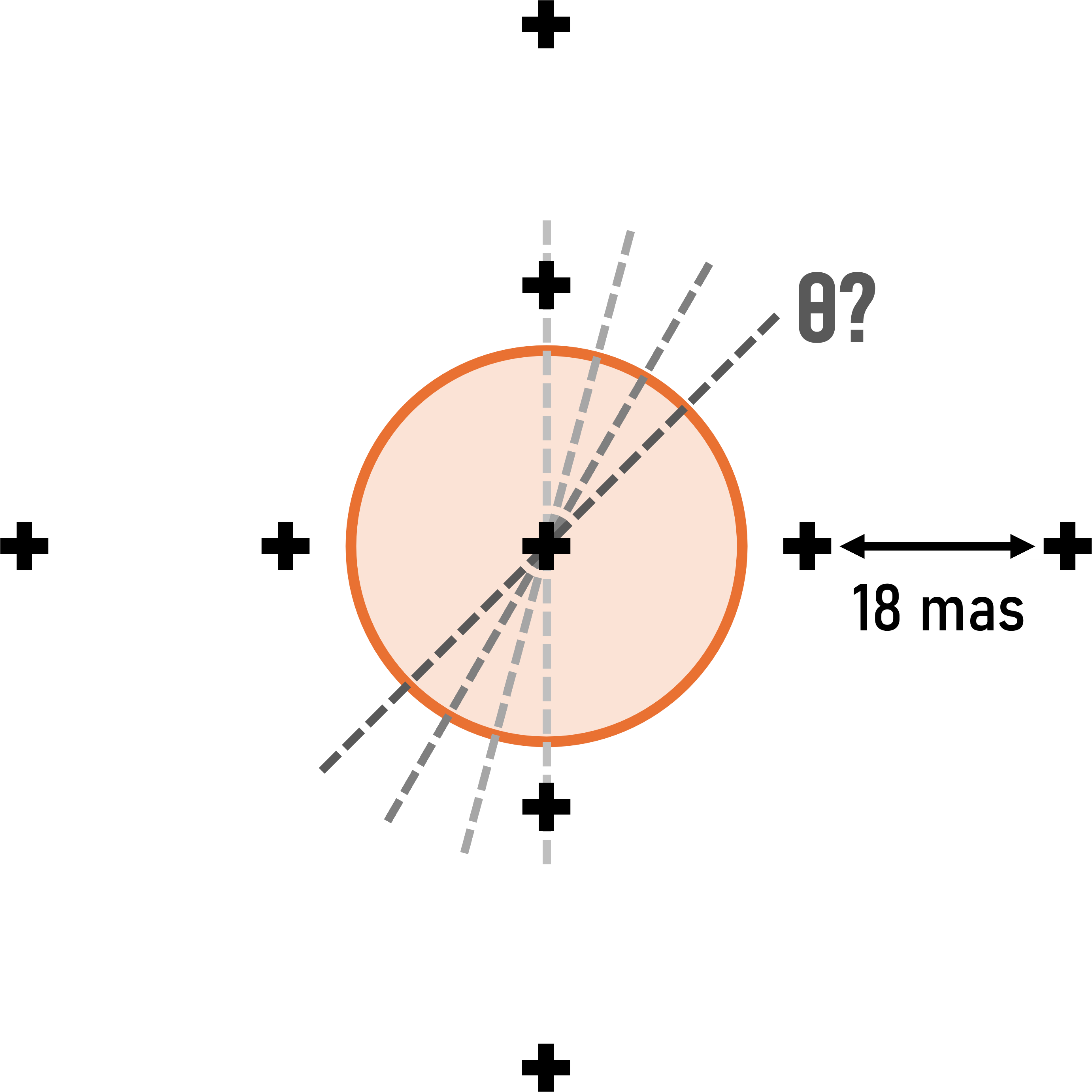}
    \vspace{0.2cm}
    \caption{Schematic of the X scan pattern employed in our DLC observations, with the positions of the fiber center relative to the star marked as crosses. The number of off-center positions can vary depending on observing conditions and available time, but the basic pattern always remains the same. Two perpendicular scans are required to ensure that at least one scan avoids the direction of minimized DLC (along the stellar spin axis). The 18~mas step size is chosen to be conveniently 0.5~pix on the PARVI guide camera.}
    \label{fig:cross_scan}
\end{figure}

We manually commanded $xy$ positions to the PARVI tip-tilt stages to offset the star along two perpendicular lines, forming an ``X'' centered on the best-guess position of the fiber center location, as shown in Fig.~\ref{fig:cross_scan}. This center position must be re-derived for each PARVI observing run (since the instrument is unplugged from the telescope between runs) by peaking up a back-injection laser with the PARVI fiber. It is necessary to scan in two perpendicular directions because the on-sky orientation of the stellar spin axis is unknown --- the RV shift from DLC is maximized when offsetting along the stellar equator and zero when offsetting directly along the spin axis, so sampling positions in both the $x$ and $y$ directions is needed to disentangle the magnitude of $\Delta$RV from the spin axis orientation.

At each position on the X, we took at least three PARVI exposures to boost S/N and help with cosmic ray rejection, and to serve as a gauge of RV scatter. We also saved a data cube consisting of 1~second of images from the tip-tilt guide camera (which generally runs at 100-400~Hz) to quantify the offset accuracy of the guide camera as well as the quality and jitter of the stellar PSF. We made sure to regularly return to the central position, where $\Delta$RV is zero if the star has been properly centered on the fiber, in order to track any bulk RV drifts over the course of the night.

We initially planned to repeat a second X pattern rotated by 45\degree\ relative to the first scan, which would give us better constraints on the spin axis orientation. However, after preliminary analysis showed that the RV behavior was far less clean than originally anticipated, we decided that it was more important to establish the repeatability of the DLC RV shifts over time, so we simply repeated a single X scan as many times as target observability and weather conditions allowed.

We calculated exposure times with a target S/N of 300 in the $J$ and $H$ spectral ranges, but we anticipated that we would have to adjust these on the fly, as past experiences have shown that PARVI flux rates can be highly variable depending on seeing and AO performance. We took all observations with simultaneous etalon light in the calibration trace on the spectrograph, for simultaneous wavelength calibration to help track and remove instrumental drift.

\section{OBSERVATIONS}
\label{sec:obs}

We observed DLC targets with PARVI on the nights of 2025 Feb~26, Apr~06, and Oct~15--16, and 2026 Apr~29, prioritizing targets with large expected $\Delta$RV and good observability. On later dates, we also prioritized re-visiting targets that we had previously observed. We were also able to get additional observations of our target stars in the normal PARVI observing configuration (i.e., at the center of the X) on several other nights to help build up their spectral templates, thanks to time-trades with other programs. We summarize these observations in Table \ref{tab:obs_summary}. The raw echellograms were processed into extracted 1D spectra with the PARVI Data Reduction Pipeline as described in Ref.~\citenum{Gibson2022JATIS...8c8006G}.

On 2025 Feb~26, we also observed visual binaries with separations of a few arcsec to validate the plate scale of the PARVI tip-tilt guide camera (Sec.~\ref{ssec:gcam_scale}), and on several epochs, we took calibration frames with a uranium-neon (UNe) atomic line lamp illuminating the science trace in order to derive an improved wavelength solution for PARVI's blue orders, as the wavelength solution previously had to be extrapolated for orders blueward of echelle 110 where the laser comb coverage ends. 

\begin{table}[tbp]
\caption{Summary of PARVI DLC observations}
\label{tab:obs_summary}
\centering
\small
\renewcommand{\arraystretch}{1.25}
\vspace{0.2cm}

\begin{threeparttable}
\begin{tabular}{lcccll}
\toprule
\textbf{Target} & \textbf{$H$ (mag)} & \textbf{$\alpha_*$ (mas)} & \textbf{$v \sin{i}$ (km/s)} & \textbf{DLC obs. dates} & \textbf{Add. on-center epochs} \\
\midrule
Mu Gem  & --1.68 & 13.9 & 8.4 & 20250226, 20251015, 20251016 & 6 \\
Zet Leo & +2.63  & 1.5  & 72  & 20250226                     & 4 \\
Del Vir & --1.01 & 9.9  & 6   & 20250226, 20250406, 20260429 & 4 \\
Alf Lyn & --0.47 & 7.2  & 6.4 & 20250406                     & 0 \\
Gam Dra & --1.03 & 10   & 6   & 20250406                     & 0 \\
Altair  & +0.24  & 3.6  & 242 & 20251016                     & 2 \\
Bet Peg & --2.13 & 16   & 9.7 & 20251016                     & 4 \\
Bet Cas & +1.43  & 2    & 71  & 20251016                     & 3 \\
\bottomrule
\end{tabular}
\end{threeparttable}
\end{table}

Since the M giants we are targeting for DLC observations are extremely bright in PARVI's $JH$ bandpass (the brightest being $H \approx -2.1$), we initially took exposures as short as 1.5~s, which is the minimum full-frame readout time for the PARVI H2RG detector. However, we discovered that this caused the spectrograph control computer to crash as a result of trying to write too many FITS files too quickly. To prevent this from happening again, we inserted fiber optic attenuators into the PARVI input fiber train to increase the exposure times to a minimum of 30~s, which has been used for extended exposure sequences in the past without issue. These attenuators provide a reliable and repeatable way to achieve 3, 5, or 10 dB of grey attenuation, and can be quickly swapped out depending on target brightness and observing conditions. Increasing the exposure times also allows for a higher S/N for the simultaneous etalon calibration source, which has a maximum flux rate --- otherwise etalon frames must be bracketed before and after the science frames, reducing observing efficiency.

While we were able to stay on-sky for the majority of our nights, the weather at Palomar can be quite temperamental, and we frequently encountered seeing poor enough that the AO struggled to produce PSFs with well-defined cores (with one recurring failure mode being an elongated, cigar-shaped PSF) and the PARVI tip-tilt system would also lose lock on the stellar PSF, either intermittently or permanently. In these situations, we would either pause observations or only take on-center exposures of the target until conditions improved again. Even when both the AO and tip-tilt systems were able to function properly, we found that the seeing, and consequently the AO performance, often varied wildly over the course of an X scan.

We also note that in between our observations on 2025 Oct~16 and additional observations of our targets taken starting Oct~28, PARVI suffered a cryocooler power outage that caused the echelle orders to visibly shift on the detector, in both the dispersion and cross-dispersion directions, once the instrument thermally re-stabilized. This required us to take a new set of fiber flats (traces) for spectral extraction and a new set of UNe exposures in order to rederive the wavelength solution. For now, we assume that any changes in the instrumental line spread function are not significant, and allow spectra from both before and after the thermal cycle to be stacked together to create our stellar templates (Sec.~\ref{ssec:serval_rvs}).

\section{ANALYSIS}
\label{sec:analysis}

\subsection{Guide camera plate scale}
\label{ssec:gcam_scale}

\begin{figure}[tbp]
    \centering
    \includegraphics[width=\linewidth]{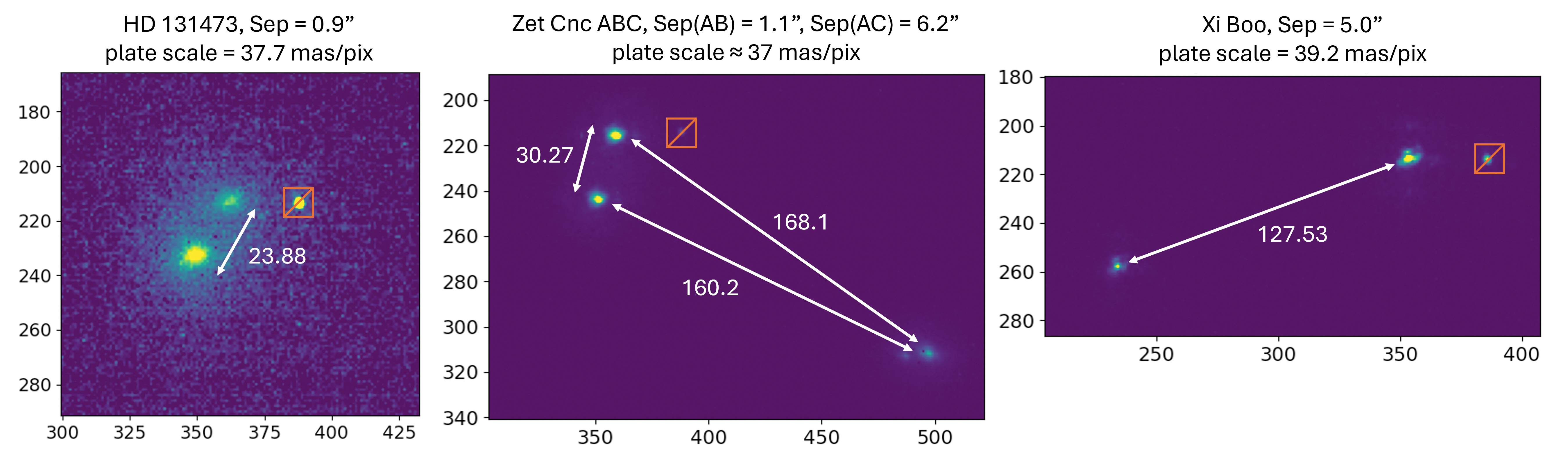}
    \caption{Three multiple star systems as imaged by the PARVI tip-tilt guide camera. Pixel distances between stellar centroids are indicated in white, and the orange box marks the spot resulting from the PARVI back-injection laser, which should be ignored in these images. All three systems yield similar plate scales of $\sim$37~mas/pix.}
    \label{fig:double_stars}
\end{figure}

In order to confirm the plate scale of the PARVI tip-tilt guide camera, we observed three available multiple star systems with known separations and proper motions from the Washington Double Star catalog\footnote{\url{https://www.astro.gsu.edu/wds/}}. We bias-subtracted and median-combined our images of the AO-corrected guide camera field, and calculated the distances between the centroids of the stellar PSFs (Fig.~\ref{fig:double_stars}). Based on this quick analysis, we derive similar plate scales for these three systems and adopt a value of 37~mas/pix, which is sufficiently precise for our purposes.

\subsection{Tip-tilt accuracy and jitter}
\label{ssec:gcam_jitter}

\begin{figure}
    \centering
    \includegraphics[width=\linewidth]{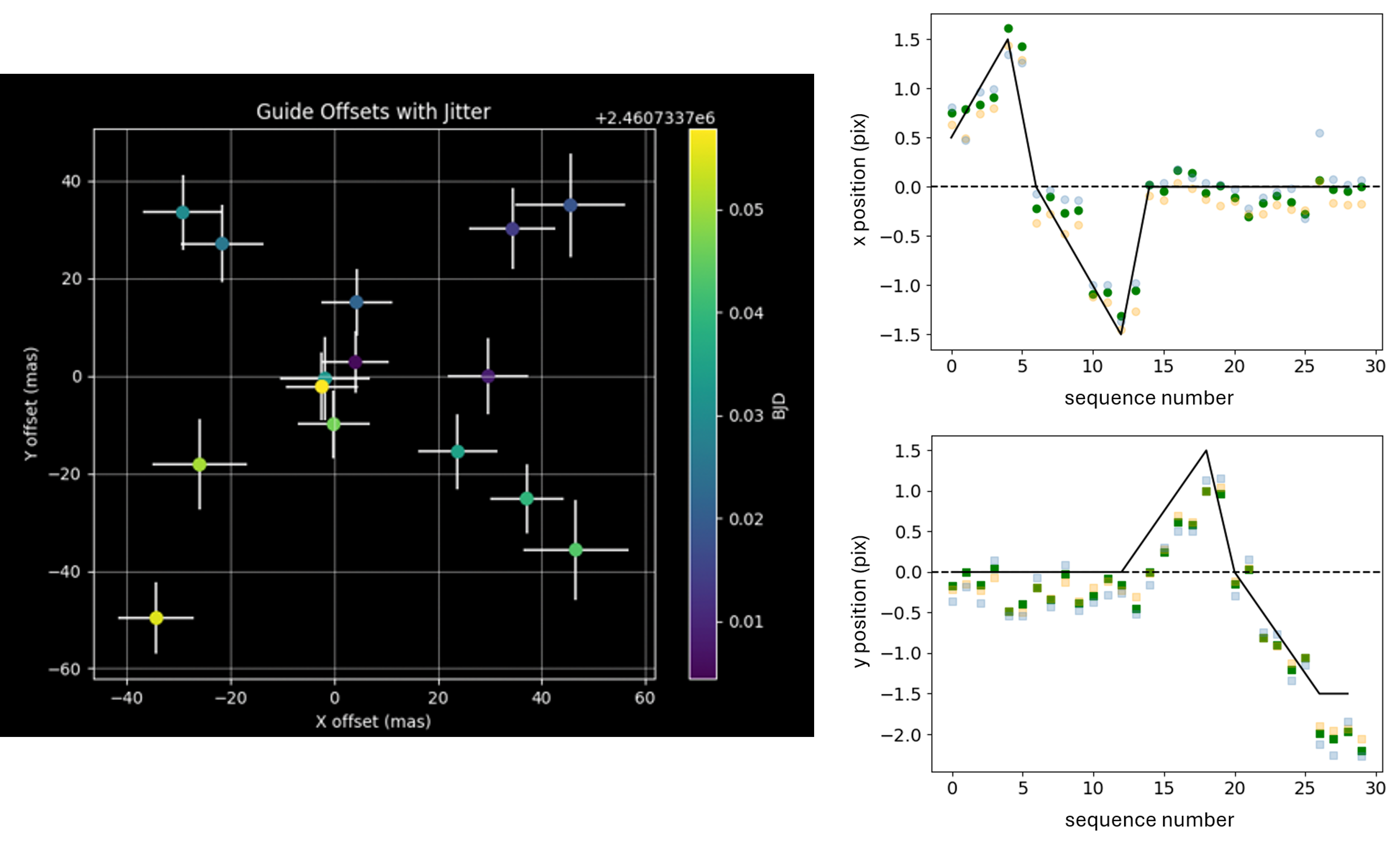}
    \caption{\textit{Left}: Map of $xy$ guide camera positions, as derived from the PSF centroid, for one set of DLC observations of Mu~Gem. While the general shape of the X scan is present, individual pointings have deviated from their commanded positions by up to 20~mas. For each position, the errorbar represents the positional jitter derived from the guide camera data cubes. \textit{Right}: Centroid-derived positions compared to commanded positions (black line) from DLC observations of Zet~Leo, showing similar levels of variation in the commanded vs. derived positions, as well as a potential systematic offset in the $y$ direction.}
    \label{fig:gcam_all}
\end{figure}

We used the image data cubes taken with the PARVI guide camera during DLC observations to quantify the accuracy of the offsets we make with the tip-tilt piezo stages. When we compare average centroid positions derived from the guide camera data to the commanded positions (Fig.~\ref{fig:gcam_all}), we see that the offsets are \textit{generally} accurate to less than our 0.5~pix step size (18~mas), but with deviations of up to 20~mas, meaning the stages are not performing as well as our expected positional accuracy of a few mas. These investigations also reveal that there may be mechanical bias in the $y$-direction piezo stage, based on a systematic offset of commanded vs. centroid-derived positions.

We were also able to use time series information from the data cubes to calculate a typical jitter for the stellar PSF of 5--10~mas (Fig.~\ref{fig:gcam_all}, left). Our predicted $\Delta$RVs from Eq.~\ref{eq:dlc} assume that the pointing offset is perfectly static, but under real observing conditions we expect PSF jitter to decrease the observed $\Delta$RV from DLC, since executing controlled jitter patterns with the tip-tilt mirror to smooth out the fiber coupling map is one of the potential DLC mitigation strategies put forth by Ref.~\citenum{Baker2024SPIE13096E..1GB}. Further work is needed to incorporate this jitter into our DLC simulations to derive a more realistic estimate for the $\Delta$RVs we can expect to see.

\subsection{Template-matching RVs}
\label{ssec:serval_rvs}

While the PARVI Data Reduction Pipeline is established as the preferred method for extracting the 1D PARVI spectra, at present there are no agreed-upon best practices for deriving RVs from those extracted spectra. 
Since many of our DLC target stars are K or M-type and large portions of PARVI's NIR bandpass are unusable for RVs due to being infested by tellurics, we thought to use the template-matching method \cite{AngladaEscude2012ApJS..200...15A} to derive RVs. In short, template-matching stacks all available spectra of the target star to produce a high-S/N template which incorporates the instrument profile, and calculates the (relative) RV of each exposure compared to this ``master'' spectrum. Because it is fitting not only the positions of spectral features, but also their shape, template-matching ought to be capable of extracting more RV information content from a given spectrum than the classic cross-correlation function (CCF) method \cite{Baranne1979VA.....23..279B}, especially for M-type stars whose spectral features are often less cleanly delineated from each other. We use the template-matching algorithm as implemented in SERVAL (SpEctrum Radial Velocity AnaLyser) \cite{Zechmeister2018A&A...609A..12Z}, based on a version of SERVAL customized for the NEID spectrograph \cite{Stefansson2022ApJ...931L..15S} which we subsequently adapted to PARVI.

Our PARVI-SERVAL RVs demonstrate very good precision, with typical errorbars of 5--10 m/s, but they also exhibit RV slopes of hundreds of m/s over the course of a night for every single one of our target stars, even after barycentric correction! Independent analyses of PARVI etalon data supported the findings of Ref.~\citenum{Cale2023JATIS...9c8006C} that PARVI's intra-night instrumental stability is $<$~5--10~m/s, and the fiber-to-fiber drift between the science and calibration fibers is an order of magnitude smaller, suggesting that the RV slopes we see in PARVI-SERVAL are not actually instrumental in nature, but possibly algorithmic.

\begin{figure}[tbp]
    \centering
    \includegraphics[width=\linewidth]{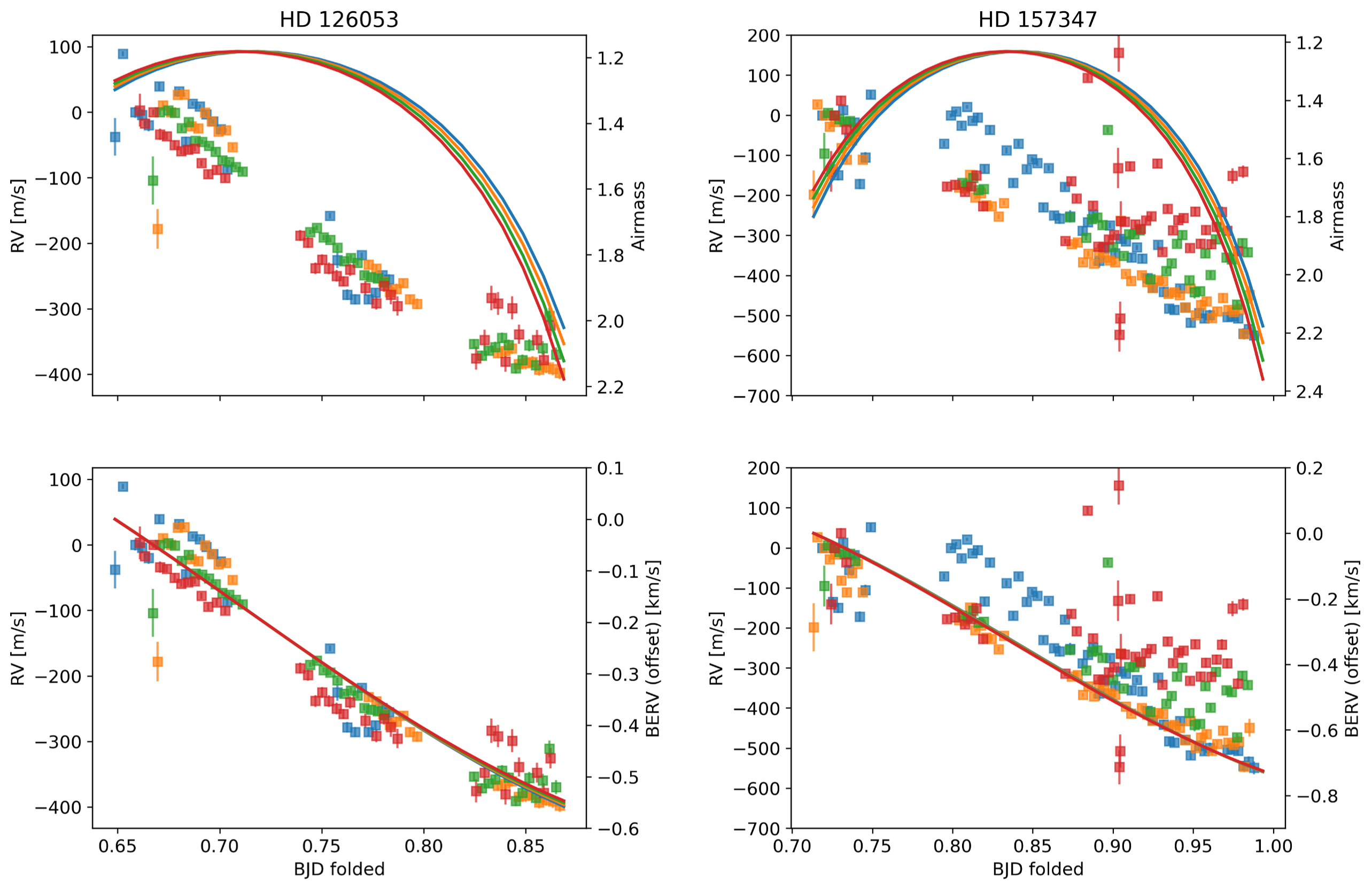}
    \caption{RVs of two EPRV standard stars derived via the template-matching method, showing a roughly linear slope of hundreds of m/s over each night that does not correlate with the airmass of the targets, but does correlate with the barycentric earth radial velocity (BERV). The zeropoints for RVs and BERVs are reset for each of the 4 different nights (different colors) to better show the common RV trend.}
    \label{fig:eprv_standards}
\end{figure}

We were able to confirm this thanks to publicly available observations of EPRV standard stars taken with PARVI. Standard stars are intended to serve as precision benchmarks for RV instruments, and are chosen based on existing observations (with some stars having nearly 30 years of RV heritage) that indicate they are either long-term RV-stable at the few m/s level or better, or less commonly, have planets with extremely well-characterized orbits. 
The standard stars HD~126053 and HD~157347 were observed with PARVI on four consecutive nights from 2025 June 5--8. Fig.~\ref{fig:eprv_standards} presents the derived PARVI-SERVAL RVs, which all show an approximately linear slope of hundreds of m/s over the course of each night. We had hypothesized that the source of the RV slope could be connected to PARVI's lack of an atmospheric dispersion corrector, but the standard star RVs clearly demonstrate that the slope is uncorrelated with target airmass. However, the RVs do show a strong correlation with the barycentric earth radial velocity (BERV) \cite{Wright2014PASP..126..838W, Kanodia2018RNAAS...2....4K}, despite the fact that PARVI-SERVAL has already corrected for the BERV shift.

We suspect that this is the same effect seen in ESPRESSO and HARPS RV data by Ref.~\citenum{Silva2025A&A...700A..93S}, who see a similar slope in template-matching and line-by-line RVs --- but not CCF RVs --- when only a few nights of data are used to construct the master template. They find that the magnitude of the RV slope decreases as the number of nights increases, suggesting that uncorrected residual effects in the observer rest frame (such as microtellurics or fixed detector pattern noise) are present in the master template but become increasingly averaged out as the number of epochs and the barycentric coverage increases, which is supported by the strong correlation between our RVs and the stellar BERVs which track the movement of the rest frame. Unfortunately, we do not see much of an improvement in the RV slope when we add more on-center data from additional nights for Mu~Gem, our DLC target with the most overall observations, though this is possibly because the number of individual exposures from the 3 DLC nights far outweighs the 6 on-center nights, and thus the extra on-center data do not have much influence on the stacked master spectrum.

Thus, we treat our template-matching RVs with considerable caution. When the intra-night slope is fit out, we see increased RV scatter in off-center vs. on-center RVs for many of our DLC targets, akin to the RV scatter of Altair observed by Ref.~\citenum{Guyon2024SPIE...1309719} with SCExAO, again showing hints of the direct RV impact of DLC. However, we also observe that this increased RV scatter is frequently associated with increased RV error bars, due to lower S/N for off-center RVs, so we refrain from drawing any definitive conclusions for the time being.

\section{CONCLUSIONS \& FUTURE WORK}
\label{sec:future}

In conclusion, we have designed and executed an observing experiment to directly measure the RV impact of DLC for the first time using the precision RV spectrograph PARVI, by deliberately using the tip-tilt system to move stars off the center of the instrument input fiber. We were able to use the data gathered to characterize the plate scale, positional accuracy, and PSF jitter of the PARVI tip-tilt guide camera. When we derive RVs from our PARVI spectra using the template-matching method, we see a consistent RV trend correlated with barycentric motion, suggesting residual observer-frame contamination in the master spectral template. We plan to explore whether this can be mitigated by modifying the PARVI-SERVAL configuration parameters, while also deriving RVs via the classic CCF method, which is largely immune to this effect. 

Deriving reliable RVs and being able to compare them against more realistic simulations of DLC (e.g., including pointing jitter) will help us improve our understanding of DLC, and move forward in developing and validating mitigation strategies on-sky. Ultimately, we hope these efforts will improve the achievable RV precision of HISPEC and aid the EPRV community in determining the feasibility of diffraction-limited spectrographs for EPRV science.

\acknowledgments 

We thank J.K.~Luhn and the Standard Stars Working Group of the EPRV Research Coordination Network for the publicly available observations of EPRV standard stars with PARVI used in this work. 

This research has made use of the Washington Double Star Catalog maintained at the U.S. Naval Observatory.

Part of this work was performed at the Jet Propulsion Laboratory, California Institute of Technology, sponsored by the United States Government under the Prime Contract 80NM0018D0004 between Caltech and NASA.

HISPEC is supported by the Caltech fund for Keck instrumentation, the Gordon-Betty Moore Foundation, the Heising-Simon Foundation, W.~M. Keck Observatory and the University of California Observatories. HISPEC benefits from in-kind contributions from NASA, the AstroBiology Center (NINS) and Northwestern University.

\bibliography{mybib} 
\bibliographystyle{spiebib} 

\end{document}